\documentclass[singlecolumn]{aastex}
\usepackage{spr-astr-addons}
\usepackage{url}

\begin{document}

\title{Late Time X-ray, IR and Radio Observations of the Tidal Disruption Event Galaxy NGC~5905}
\shorttitle{Tidal Disruption Galaxy NGC~5905}
\shortauthors{Raichur et al.}

\author{H. Raichur\altaffilmark{1,2}} 
\affil{Raman Research Institute, Bangalore, India\altaffilmark{1}}
\affil{NORDITA, KTH Royal Institute of Technology and Stockholm University, 
Roslagstullsbacken 23, SE-10691 Stockholm, Sweden\altaffilmark{2}}
\and 
\author{M. Das\altaffilmark{3}}
\affil{Indian Institute of Astrophysics, Koramangala, Bangalore 560034, India\altaffilmark{3}}
\and
\author{A.Alonso Herrero\altaffilmark{4}}
\affil{Instituto de Fisica de Cantabria, CSIC-UC, Avenida de los Castros S/N, ES 39005 Santander, Spain\altaffilmark{5}}
\and
\author{P.Shastri\altaffilmark{3}}
\affil{Indian Institute of Astrophysics, Koramangala, Bangalore 560034, India\altaffilmark{3}}
\and
\author{N.G.Kantharia\altaffilmark{5}}
\affil{National Centre for Radio Astrophysics, TIFR, Pune, India\altaffilmark{6}}

\begin{abstract}
NGC~5905 is one of the few galaxies with no prior evidence for an  
active galactic nucleus (AGN) in which an X-ray flare was detected by the 
{\it ROSAT} All Sky Survey (RASS) in 1990-91. Follow-up studies showed that the 
X-ray flare was due to the tidal disruption of a star by the massive black 
hole in the center of the galaxy. In this study we present analysis of 
late-time follow-up observations of NGC~5905 using {\it Chandra} archival data, 
{\it Spitzer} archival data, GMRT 1.28~GHz radio observations and VLA 3~GHz and 8~GHz archival 
data. The X-ray image shows no compact source that could be associated 
with an AGN. Instead, the emission is extended -- likely due to nuclear star 
formation. The total measured X-ray luminosity from this extended emission 
region is comparable to the X-ray luminosity determined from the 2002 
{\it Chandra} observations and is a factor of 200 less than the peak of the X-ray 
flare observed in 1990. Diffuse X-ray emission was detected close to the 
circum-nuclear star forming ring. The {\it Spitzer} 2006 mid-infrared spectrum 
also shows strong evidence of nuclear star formation but no clear AGN signatures. 
The semi-analytical models of \cite{tommasin.et.al.2010} together
with the measured [OIV]/[NeII] line ratio suggest that at most only 5.6\% of 
the total IR Flux at 19~$\mu$m is being contributed by the AGN. 
The GMRT 1.28~GHz observations reveal a nuclear source. 
In the much higher resolution VLA 3~GHz map, the 
emission has a double lobed structure of size 2.7$^{\prime\prime}$
due to the circumnuclear star forming ring. The GMRT 1.28~GHz peak emission coincides 
with the center of the circumnuclear ring. 

We did not detect any emission in the VLA 8~GHz (1996) archival data. 
Instead we give upper limits to the radio afterglow of the tidal disruption event (TDE) using
3~$\sigma$ upper limits where $\sigma$ is the map
noise. The 3~$\sigma$ limits at 1.28~GHz, 3~GHz and 8~GHz are 0.17~mJy, 0.09~mJy and 0.09~mJy, respectively.
Our studies thus show that (i)~NGC~5905 has a declining X-ray flux consistent 
with a TDE, (ii)~the IR flux is dominated by nuclear star formation, 
(iii)~the nuclear radio emission observed from the galaxy is due to 
circumnuclear star formation, (iv)~no compact radio emission associated with a radio 
afterglow from the TDE is detected.
\end{abstract}

\keywords{Galaxies: spiral - Galaxies: individual (NGC~5905) - Galaxies: nuclei - Galaxies: X-ray
- Galaxies: active - X-rays - radio continuum - infrared radiation.}

\section{Introduction}
Tidal disruption of stars at the centers of galaxies are indicators for the presence of supermassive black holes (SMBHs)
in their nuclei.
Theoretical studies show that when a star passes close to a SMBH it can be tidally disrupted. The 
tidal disruption event (TDE) results in about half of the stellar mass falling into the black hole, while the
remaining tidal debris is ejected out from its orbit \citep{rees.1988, ulmer.1999, strubbe.quataert.2009}. This 
tidal debris will fall back onto the accretion disk and result in enhanced accretion rates and outflows that
can persist over much longer timescales than the actual TDE (see Bower et al. 2013 and references therein). 
Radio jets can be triggered by the infalling debris but the timescale depends on the accretion rates.
In some cases they can be triggered several years after the TDE \citep{Bow11} and will appear as compact emission
associated with the nucleus. In this paper, we analyze late time multi-wavelength observations of the TDE in 
NGC~5905 with a special focus on searching for radio jets that may be triggered by the TDE. Late time radio 
emission can arise from within the jet itself
\citep{vanvelzen.etal.2011} or from the interaction of the radio jet with the circumnuclear medium
\citep{giannios.metzger.2011}. The resulting blast wave expands into the surrounding circumnuclear gas and later
enters a slower Sedov-Taylor expansion phase which can be detected in radio emission several years after the TDE.
Since most of the TDE host galaxies are at distances of several Mpc \citep{ESKD12, Sax12, cappelluti.etal.2009, 
gezari.etal.2009, gezari.etal.2008}, the radio jets will appear as compact nuclear emission. To date only a few TDEs have
been detected in radio emission. Two recent TDEs that were detected in $\gamma$ and X-ray emission showed variable
radio emission that decayed over a timescale of months \citep{bloom.etal.2011, zauderer.etal.2011}. Another source
IC~3599, that was detected in X-ray by {\it ROSAT} showed, however, late time radio emission \citep{Bow13}. This
indicates that late time radio jets or a radio afterglow due to jet-cloud interaction can be observed in TDEs.
Bower et al. (2013) also included NGC~5905 in their study but they did not detect any compact radio source.

\begin{figure}
\includegraphics[scale=0.35,angle=-90]{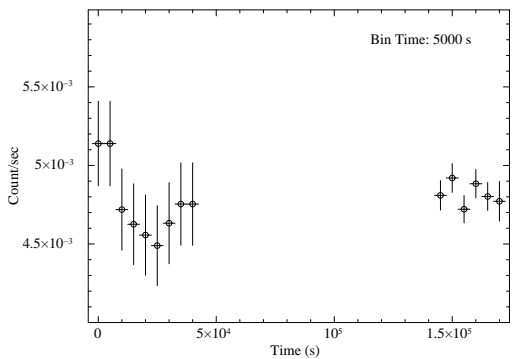}
\caption{The first and second segment of the above X-ray light curve were extracted using {\it Chandra} data from Obs-Id~7728 
and Obs-Id~8558, respectively. There is no detectable variability
during these observations when the dithering effect is taken into account.}
\label{NGC5905_X-ray_lc}
\end{figure}

\begin{table}
\caption{Details of Observational Data}
\resizebox{\columnwidth}{!}{%
\begin{tabular}{lcccc}
\hline
Wave-& Instrument & Date of Observation & Exposure,Obs-Id \\
length&            &                     & time \\
\hline
X-ray & ACIS-S & 2007-06-07 & 44.95~ks (7728)\\  
      &        & 2007-06-09 & 26.95~ks (8558)\\ 
Radio & GMRT 1.28~GHz & 2011-06-10 & ~2 hours \\
             & VLA FIRST     & May 1997       &  ....    \\
             & VLA 3~GHz     & 2012-06-04 & ~15 min  \\
             & VLA 8~GHz (archive) & 1996-11-03 & ~45 min      \\
 IR    & {\it Spitzer}/IRS & \small 2006-01-16 & SH 30s $\times$ 2 cycles\\ 
             &                    &                            & LH 60s $\times$ 2 cycles \\
\hline
\end{tabular}
}
\label{obs_detail}
\end{table}

\begin{figure*}
\begin{minipage}{160mm}
\centering
\includegraphics[scale=0.5, angle=-90]{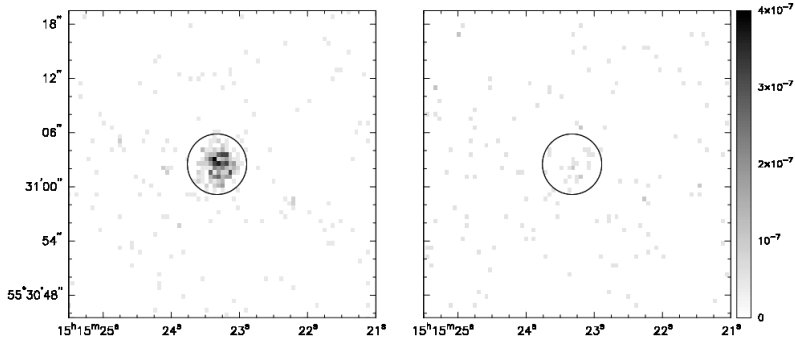}
\caption{{\it Chandra} image of NGC~5905 in the energy band 0.5-2.0 keV (left panel)
and 2.0-7.0 keV (right panel) is shown. The grey scale numbers are in the units photons~cm$^{-2}$~s$^{-1}$.
The circle of 3.5$^{\prime\prime}$ radius marks the nuclear region of NGC~5905 from which the X-ray spectrum is extracted.}
\label{X-ray_image}
\end{minipage}
\end{figure*}

{\it ROSAT} All-sky survey (RASS) discovered an X-ray flare in 
NGC~5905 in 1990. During the X-ray flare the soft X-ray luminosity reached 
a peak of $L_X~\sim~6~\times~10^{41}$~ergs~s$^{-1}$ (assuming an absorbed powerlaw 
spectral model with $N_H = 1.5 \times 10^{20}$ cm$^{-2}$, a power law index of 4 and distance 47.8 pc)
and decreased below $9 \times 10^{40}$ ergs s$^{-1}$ within 5 months of the 
X-ray outburst \citep{BKD96}. Very few cases of X-ray outbursts of
variability amplitude greater than 80 have been observed 
\citep{donley.etal.2002, esquej.etal.2007, saxton.etal.2012, maksym.etal.2010}.
Follow-up X-ray observations using {\it ROSAT} and {\it Chandra} show that the X-ray 
luminosity of the galaxy declined as $L_X~\propto~t^{-5/3}$, as expected
in the event of tidal disruption of a star near the central BH 
\citep{rees.1988, komossa.bade.1999, halpern.gazari.komossa.2004}.
Using the stellar velocity dispersion in the bulge of NGC~5905 
\citep[$\sigma = 174.6 \pm 9.0 \rm {km/s}$,][]{Ho.greene.Filippenko.2009}
and the $M_{BH}-\sigma$ calibration \citep{Gultekin.et.al.2009}, the
calculated mass of the nuclear BH in NGC~5905 is $M_{BH} \sim 10^{7}~M_{\odot}$.
This is consistent with the upper limit of $M_{BH} = 10^8~M _{\odot}$
derived by assuming that the 1990 X-ray flare was due to tidal 
disruption of a star \citep{halpern.gazari.komossa.2004}.  
Hence the nucleus of NGC~5905 may have a SMBH of mass $\sim 10^{7}~M_{\odot}$.

NGC~5905 is a barred spiral galaxy (SB(r)b type) at a distance
of 47.8 Mpc (z=0.011 and H$_0$=73 km s$^{-1}$ Mpc$^{-1}$). It
is classified as a giant low surface brightness (GLSB) galaxy because of its low
surface brightness optical disk (\citealt{kent.1985}; \citealt{sprayberry.etal.1995}).
Like most GLSB galaxies, it is relatively poor in star formation \citep{vanMoorsel.1982}
but it does have a bright bulge and nuclear star formation. H$\alpha$ observations reveal a
circum-nuclear ring of star formation of $\approx$~1~kpc in diameter \citep{mazucca.etal.2008,Com10}.
Ground based optical observations of NGC~5905 classified the nucleus as an HII  
or starburst type nucleus (\citealt{G91}, \citealt{Ho95}). However, post-outburst 
Hubble Space Telescope/Space Telescope Imaging Spectroscopy (HST/STIS) observations 
place it in the low luminosity Seyfert 2 category depending on
the estimated line ratios \citep{gezari.etal.2003}. These observed lines could not be explained 
as arising due to photoionization of clouds surrounding the central BH illuminated by the soft X-ray 
flare observed in 1990. But at the same time no direct evidence for a non-stellar continuum 
was found that could explain the observed line ratios. Thus it is not clear whether there is 
an AGN in NGC~5905 but weak AGN activity 
cannot be ruled out. Although, in general, AGNs are rare in GLSB galaxies and the BH masses are 
lower than those observed in normal galaxies \citep{ramya.etal.2011,Naik10}.

In this study we take a closer look at the late time nuclear emission from NGC~5905. We first examine the 2007 X-ray 
observations to constrain the decay of the X-ray flare. We then examine the {\it Spitzer}, {\it Chandra} and radio data to see
if there is a weak AGN in the galaxy (both the {\it Chandra} and {\it Spitzer} data, though archival, are previously unpublished).
Finally we investigate the radio emission using both the VLA 3~GHz data \citep[analyzed as part of][]{Bow13}
as well as low frequency GMRT 1.28~GHz data.  We compare the
3~GHz and 1.28~GHz maps at different resolutions to see if there is a radio jet/afterglow associated with the TDE.
We have also re-analyzed the VLA 8~GHz
archival data of 1996.
Throughout the paper luminosity of the source is estimated assuming 
$z=0.0113$ for NGC~5905 and $H_0 = 73~{\rm km~s^{-1}~Mpc^{-1}}$.

\section{Observations and Analysis}
\subsection{{\it Chandra} X-ray observations}

NGC~5905 was observed using the Advanced CCD Imaging Spectrometer (ACIS)  
on board the {\it Chandra} X-ray telescope 
twice in 2007. Table \ref{obs_detail} gives the details of the observations. 
Sometimes large flares are detected in the ACIS 
background light curve and if such flares are detected during the observation then
data recorded during the background flares should be 
filtered\footnote{{\it Chandra} memo for ACIS background flares}. 
No background flares were detected in the ACIS background light curves extracted from 
the 2007 observations.  

{\it Chandra} 2007 data was used to extract (a)~the X-ray light curve of NGC~5905,
(b)~the image of NGC~5905 in two X-ray energy bands, (c)~the X-ray spectrum 
of the central region and (d)~the image of diffuse X-ray emission.
We first examined the source light curve for X-ray flux variability
using the {\it Chandra} task {\texttt{glvary}}. The source light curve
extracted using {\texttt{glvary}} takes into account the variability due to dithering
and optimizes the bin time for the light curve accordingly. A quick look at the light curve
in Figure \ref{NGC5905_X-ray_lc} shows no count rate variability. From the results of
{\texttt{glvary}} we find that the probability that the observed signal is variable is
0.2 and the variability index is 0 implying that no variability was observed during 
the 2007 observations of NGC~5905.
\begin{table*}
\centering
\begin{minipage}{155mm}
\caption{Parameters of the X-ray spectrum model}
\resizebox{1.0\columnwidth}{!}{%
\begin{tabular}{lccccccccccc}
\hline
 & & \multicolumn{2}{c}{Component I}&\multicolumn{2}{c}{Component II}& & \\
Model&${\rm N_H}$&$\Gamma$&Norm&kT&Norm&$\chi^2$&dof&$L_{{\rm{x}}(0.5-2~{\rm{keV}})}$&$L_{{\rm{x}}(2-7~{\rm{keV}})}$ \\
&($10^{22}\rm{cm}^{-2}$)& &photons cm$^{-2}$keV$^{-1}$&keV& & & &$\rm{erg~s}^{-1}$&$\rm{erg~s}^{-1}$\\
\hline
wabs(raymond)&$0.015${(fixed)}&--&--&$0.76^{+0.06}_{-0.08}$&$(2.5 \pm 0.6) \times 10^{-5}$&
$13.79$&$13$&$(3.2 \pm 0.2)\times10^{39}$&$(2.5\pm0.4)\times 10^{38}$\\ 
wabs(pwlw+bb)$^a$&$0.71^{0.48}_{-0.38}$&$3.29^{2.01}_{-1.85}$&$7.34^{2.38}_{-6.19}\times10^{-6}$
&$8.53^{0.04}_{-0.02}\times10^{-2}$&$2.96^{264}_{-2.82}\times10^{-5}$&$12.79$&$11$& 
$(3.2\pm0.3)\times10^{39}$&$(7.1\pm4.8)\times10^{38}$\\
wabs(pwlw)$^a$&$0.67^{0.36}_{-0.25}$&$6.79^{2.47}_{-1.71}$&$3.87^{6.44}_{-2.05}\times10^{-5}$&
--&--&$25.74$&$13$& $(3.1 \pm 0.3) \times 10^{39}$ & $(1.1 \pm 0.1) \times 10^{37}$\\  
wabs(bb)$^b$&$0.17$& -- & -- &$0.17$&$4.33\times10^{-7}$&$31.44$&$13$ & 
$3.11 \times 10^{39}$ & $2.48 \times 10^{37}$ \\
wabs(bb)$^b$&$0.015$ {(fixed)}& -- & -- &$0.22$&
$1.90\times10^{-7}$&$35.41$&$14$ &
$3.06 \times 10^{39}$ & $7.81 \times 10^{37}$ \\
\hline
\end{tabular}
}
\\$^a${Fixing the $\rm{N_H}$ value to the Galactic ${\rm N_H}=1.5\times10^{20}$cm$^{-2}$ gives a bad fit with $\chi^2_{\nu} > 2$}\\ 
$^b${Reduced $\chi^2$ is greater than 2 and hence no uncertainties on the fitted parameters can be calculated}\\
\label{X-ray_spec_model}
\end{minipage}
\end{table*}

\begin{figure}
\centering
\includegraphics[scale=0.35, angle=-90]{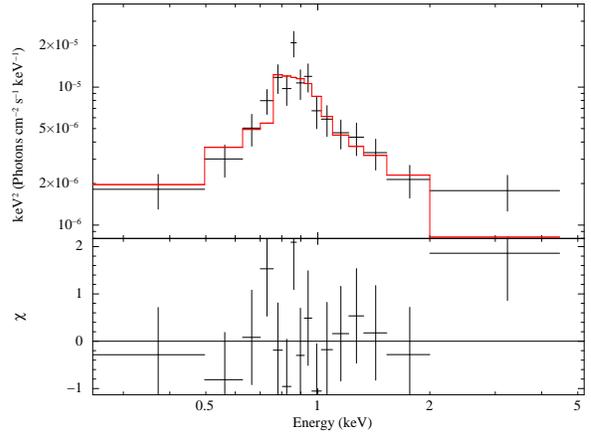}
\caption{The {\it Chandra} X-ray spectrum of NGC~5905 extracted from the region marked 
in figure \ref{X-ray_image} along with the best-fit model (red curve) and the $\chi$ 
residuals (lower panel). The model is calculated assuming absorbed emission from a hot diffuse gas. 
See Table \ref{X-ray_spec_model} for
the model details.} 
\label{X-ray_spec}
\end{figure}
Since the source flux did not vary during the two observations of 2007, 
we merged the events obtained from both the pointings and made a combined image of NGC~5905 with better
statistics in two energy bands namely 0.5-2.0 keV and 2.0-7.0 keV. Before merging the events,
we matched the aspect solution of 
Obs-Id~8558 to that of Obs-Id~7728 using an X-ray point source in the field of view
located at RA~15h~15m~18.254s and Dec~+55$^{\circ}$~32$^{\prime}$~53.71$^{\prime\prime}$. 
Since both the observations have
similar pointing offsets of 0.005', the shift is very small, 
$\Delta \rm{RA} = {8.63 \times 10^{-6}}{\rm deg}$ and $\Delta \rm{Dec} = -{1.52 \times 10^{-5}}{\rm deg}$, 
corresponding to a physical shift in x-direction of -0.06 pixels and in y-direction of 
-0.11 pixel. After reprojecting the events to a different tangent point
using the matched aspect solution, combined images in the two energy bands 
were extracted using the task {\texttt{fluximage}}. Figure \ref{X-ray_image} 
shows the exposure corrected image of NGC~5905 in the two energy bands. 
No central peak near or at the location of the optical center is seen in either of the X-ray images. 
The observed image of the nuclear region of NGC~5905 is a cluster of bright pixels at the 
0.5-2.0 keV energy range. On the other hand NGC~5905 is hardly discernible above the background in the 
(2.0-7.0) keV energy image. The number of counts within the 3.5$^{\prime\prime}$ radius nuclear region marked in Figure 
\ref{X-ray_image} is 213 and 30 in the 0.5-2.0 keV and 2.0-7.0 keV energy bands, respectively. This  
gives a hardness ratio of 0.7 for the nuclear region of NGC~5905.

The spectrum of NGC~5905 is extracted from a circular region of 3.5$^{\prime\prime}$ radius, shown in Figure \ref{X-ray_image}.
We extracted the spectra and responses separately for the two pointings and then co-added them
to get a final combined spectrum. Corresponding background spectra were extracted from a source free region on the CCDs.
The spectra and corresponding responses were extracted using the 
task {\texttt{specextract}} and coadded using the task {\texttt{combine\_spectra}}. We grouped the 
combined spectrum such that each energy bin has 15 counts. Figure \ref{X-ray_spec} shows the spectrum
of NGC~5905 with the best fit model which describes the observed X-ray spectrum as emission due to hot, diffuse gas. 
\begin{figure}
\centering
\includegraphics[scale=0.4,angle=-90]{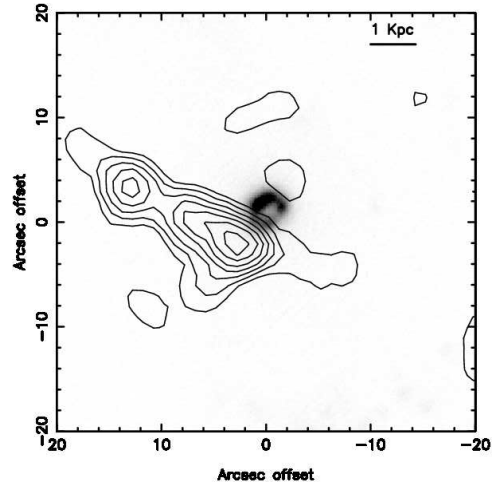}
\caption{The H$\alpha$ map taken from the AINUR atlas is plotted in gray scale.
Overlaid on this map are the contours of the X-ray diffuse emission smoothed over 4 pixels. 
The contours are 5$\sigma$ to 11$\sigma$ times the noise level in steps of 1$\sigma$. } 
\label{X-ray_diffuse_emission}
\end{figure}
To derive the diffuse emission in NGC~5905 we followed the procedure
used in \cite{das.et.al.2009}. The field of view of CCD-Id:~7 was used
and only events in the energy band of 0.5-7.0~keV were included. First, sources
in the field of view were detected using {\texttt{wavdetect}} and then separate
source and background region files for the detected sources were created 
using {\texttt{roi}} and {\texttt{splitroi}}. The source count rates were 
replaced by corresponding background counts for each source region and an image 
for each pointing was created by {\texttt{dmfilth}}. Images and exposure maps,
thus created for each pointing, were combined using the tasks 
{\texttt{reproject\_image\_grid}} and {\texttt{reproject\_image}}. 
The combined diffuse emission image was first exposure-corrected and then divided by the
exposure map to produce the required flux image. We detected weak, diffuse 
emission near the nucleus of NGC~5905. The peak emission has a value approximately
14 times the background emission and this emission lies close to the circumnuclear ring of star formation,
as shown in Figure~\ref{X-ray_diffuse_emission}, where the contours of the diffuse emission 
are overlaid on the H$\alpha$ image which is obtained from the Atlas of Images of
NUclear Rings \citep[AINUR,][]{Com10}. 
\begin{figure}[t]
\includegraphics[scale=0.4,angle=-90]{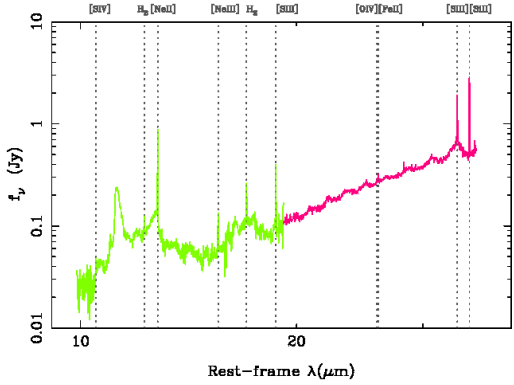}
\caption{{\it Spitzer}/IRS high spectral resolution (SH+LL) spectrum of NGC~5905. 
The position of the most prominent emission lines are marked (see Table \ref{IR-line-flux} and text).
}
\label{spitzer_spectrum}
\end{figure}
\begin{table}
\caption{Line fluxes of IR spectral lines}
\resizebox{0.8\columnwidth}{!}{%
\begin{tabular}{lcc}
\hline
Line & Wavelength & Fluxes  \\
element & ($\mu\,\rm{m}$) & ($10^{-14}\rm{ergs\;s^{-1}\;cm^{-2}}$) \\
${\rm [PAH]}$      & 11.25 & $142.00 \pm 1.42$\\
${\rm [Ne II]}$    & 12.82 & $21.64 \pm 3.18$ \\
${\rm [Ne III]}$   & 15.55 & $2.75 \pm 0.39$ \\
${\rm [S III]}$ & 18.71 & $8.49 \pm 1.32$ \\
${\rm [O IV]}$     & 25.89 & $0.57 \pm 0.32$ \\
${\rm [Fe II]}$    & 24.52 & $1.15 \pm 0.34$ \\
${\rm [S III]}$ & 33.48 & $21.06 \pm 2.55$ \\
${\rm [Si II]}$    & 34.8 & $38.16 \pm 3.38$ \\
\hline
\end{tabular}
}
\label{IR-line-flux}
\end{table}

\subsection{{\it Spitzer} mid-infrared observation}
The mid-infrared data were retrieved from the {\it Spitzer} archival
spectroscopic observations (Program ID: IG-A02120140, PI: A. Zezas)
obtained with the infrared spectrograph \citep[IRS,][]{houck.2004}.
The observations were taken with the short-high (SH) and long-high
(LH) modules covering the spectral ranges $9.9-19.6\,\mu$m and
$18.7-37.2\,\mu$m, respectively. The slit width was $4.6^{\prime\prime}$ for SH and $9.0^{\prime\prime}$ for LH.
The observations were reduced following the procedure described in detail by \cite{pereira-santaella.2010}.

Several fine structure lines are observed in the 
spectrum along with the $11.3\,\mu{\rm m}$ polycyclic aromatic hydrocarbon (PAH) feature (see Fig. \ref{spitzer_spectrum}). The measured
line fluxes are given in Table \ref{IR-line-flux}. To measure the flux and 
equivalent width (EW) of the 11.3$\mu$m PAH feature we used the method of 
\cite{hernan-caballero.2011}, which models the feature using a Lorentz function. 
There is no indication of the presence of the [NeV] lines
at $14.3\,\mu$m and $24.3\,\mu$m and we put upper limits on their
line fluxes at $0.54 \times 10^{-14}\,\rm{ergs\;s^{-1}\;cm^{-2}}$ and 
$0.35 \times 10 ^{-14}\,\rm{ergs\;s^{-1}\;cm^{-2}}$, respectively.
The measured value of the equivalent width (EW) of the $11.3\,\mu{\rm m}$ PAH feature is $1.28 \pm 0.02\,\mu$m
which is typical of high metallicity star-forming galaxies \citep[see][]{hernan-caballero.2011}.

\subsection{Radio Data}
\begin{figure}
\centering
\includegraphics[scale=0.4,angle=-90]{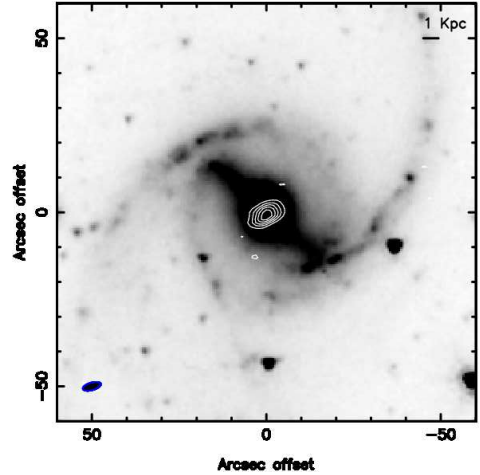}
\caption{The figure shows the contours of radio continuum emission at 1.28~GHz 
overlaid on the 3.6$\mu\,\rm{m}$ IRAC near-infrared 
image of the galaxy. The contours are 5,10,15,20 and 25 times the noise level, 
which is 0.12~mJy/beam. The beam is elliptical $5.38^{\prime\prime}\times2.21^{\prime\prime}$
and is shown in the bottom left corner. The 
radio emission is concentrated about the nucleus within the small but bright 
bulge in the galaxy.} 
\label{radio_emission}
\end{figure}
\begin{table}[t]
\caption{\bf{Radio Fluxes at different frequencies (with Robust=0)}}
\resizebox{0.9\columnwidth}{!}{%
\begin{tabular}{lccccccc}
\hline
\small $\nu$ & \small Beam  & \small Peak   & \small Noise       & \small Total & \small Fitted  \\
\small (GHz) &              & \small Flux   & \small (mJy/b)  & \small Flux     & \small Size  \\
             &              & \small (mJy/b) &             & (mJy) & \\
\hline
\small 1.28 & \small $5.4^{\arcsec}\times2.2^{\arcsec}$ & 3.84 & \small 0.12 & \small 9.26 & $6.8^{\arcsec}\times4.2^{\arcsec}$ \\
\small 1.42 & \small $5.4^{\arcsec}\times5.4^{\arcsec}$ & 6.44 & \small 0.15 & \small 9.48 & $10.0^{\arcsec}\times7.9^{\arcsec}$ \\
\small 3.0  & \small $2.1^{\arcsec}\times1.4^{\arcsec}$ & 1.04 & \small 0.03 & \small 6.18 & $5.1^{\arcsec}\times3.5 ^{\arcsec}$\\ 
\small 8.0  & \small $0.26^{\arcsec}\times0.19^{\arcsec}$ & ... & \small \bf 0.030  & \small ...  & ... \\ 
\hline
\small 1.28$^a$ & \small $4.6^{\arcsec}\times1.7^{\arcsec}$ & 1.32 & \small 0.15 & \small 3.07 & $5.5^{\arcsec}\times2.7^{\arcsec}$\\ 
\small 1.28$^b$ & \small $11.7^{\arcsec}\times3.8^{\arcsec}$ & ... & \small 0.20 & \small ... \\
\small 3.0$^a$  & \small $1.8^{\arcsec}\times1.2^{\arcsec}$ & 0.42\dag & \small 0.02 & \small 1.87 & ...  \\  
\small 3.0$^b$  & \small $1.7^{\arcsec}\times1.0^{\arcsec}$ & 0.18\ddag & \small 0.03 & \small 0.23 & ...  \\
\hline 
\end{tabular}
}
~\\
$^a$ uv cutoff 20~k$\lambda$ ~~~~~~~~~~~ \dag left lobe\\
$^b$ uv cutoff 40~k$\lambda$ ~~~~~~~~~~~ \ddag right lobe
\label{radio_obs_detail}
\end{table}
{\bf(i)~1.28~GHz~:~}NGC~5905 was observed during June 2011 in radio continuum 
at 1.28~GHz using the Giant Metrewave Radio Telescope (GMRT) located 
near Pune, India \citep{ananthakrishnan.rao.2001}.
Nearby radio sources, 1438+621 and 3C~286, were used for 
phase and flux calibrations, respectively. NGC~5905 had a two hour scan. The data were
analyzed using AIPS\footnote{Astronomical Image Processing 
System (AIPS) is distributed by NRAO which is a facility of NSF 
and operated under cooperative agreement by Associated Universities, Inc.} \citep{Greisen03}
and iteratively edited and calibrated until satisfactory gain 
solutions were obtained. Low resolution (robust 5) and
high resolution (robust 0) maps were made using IMAGR. 
Lower end UV cutoffs of 20~k$\lambda$ and 40~k$\lambda$ (i.e. baselines smaller than 20 or 40~k$\lambda$
were omitted) were applied to see if there was any compact nuclear emission. \\
{\bf(ii)~1.435~GHz~:~}We used the VLA Faint Images of the Radio Sky at Twenty-Centimeters 
(FIRST) map\footnote{Faint Images of 
the Radio Sky at Twenty-Centimeters} \citep{Becker.et.al.1995} to determine flux at this frequency. The observations were 
from the epoch May, 1997. \\
{\bf(iii)~3~GHz~:~} Observations were done in B configuration using the VLA in June 2012. 
For further details see Bower et al. (2013). We made robust 0 and robust 5 maps. Lower end UV cutoffs of
20~k$\lambda$ and 40~k$\lambda$ were applied to obtain higher resolution images and to check if there was
any compact nuclear emission.  \\
{\bf(iv)~8~GHz~:~} We used 8~GHz radio data from the VLA archive; the observations 
were from the epoch November, 1996 and were taken in the A configuration. We analyzed 
the data using AIPS; 1331+305 and 1510+570 were used as flux and phase calibrators, respectively. 
The map noise was 29.9$\mu$~Jy/beam, where the beam is 0.28$^{\prime\prime} \times 0.19^{\prime\prime}$,
and is consistent with the earlier analysis of \cite{komossa.dahlem.2001}. We did not detect any compact emission
in either robust=0 or robust=-5 (highest resolution) maps. The 5$\sigma$ upper limit to any radio
afterglow is thus 150$\mu$~Jy and is consistent with the $5\sigma$ upper limit given by
\cite{komossa.dahlem.2001}

\section{Results}
\subsection{Long term behavior of X-Ray Luminosity of NGC~5905}
The best fit model (emission due to hot diffuse gas) to the 2007 X-ray spectrum yields  
the X-ray flux of the source as $f_{\rm{x}}=(1.5 \pm 0.1)\times10^{-14} \rm{erg~s}^{-1}\rm{cm}^{-2}$
in the energy band of $0.3-8.0~\rm{keV}$ (also see Table \ref{X-ray_spec_model}). 
We also tried other models, the details of
which are presented in Table \ref{X-ray_spec_model}. The powerlaw model with variable
$N_H$ gives a very high value for the powerlaw index. A purely powerlaw model with 
absorption column density fixed to galactic value ($N_H=1.5\times10^{20} \rm{cm}^{-2}$)
gives a very poor fit. We also tried an absorbed blackbody model with variable and fixed $N_H$
values but both the models give poor fits. We also tried to model the spectrum as  
thermal bremsstrahlung but the resultant fit was very poor. For comparison, 
the 0.1-2.4 keV X-ray luminosity of NGC~5905 during the 2007 
{\it Chandra} observations is a factor of $\sim 200$ less than the luminosity 
observed during the peak of the X-ray flare in 1990.

\begin{table}
\caption{X-ray Luminosity}
\resizebox{0.8\columnwidth}{!}{%
 \begin{tabular}{cc}
 \hline
 \small Time & \small $L_{\rm x}$ (0.1-2.4)keV \\
 \small (year) & \small ($\rm{erg~s^{-1}}$) \\
 \hline
 1990.53$^a$ & $(6.45 \pm 2.29) \times 10^{41}$ \\
 1990.54$^a$ & $(8.05 \pm 2.05) \times 10^{41}$ \\
 1990.54$^a$ & $(1.17 \pm 0.18) \times 10^{42}$ \\
 1990.54$^a$ & $(1.28 \pm 0.21) \times 10^{42}$ \\
 1990.54$^a$ & $(1.17 \pm 0.17) \times 10^{42}$ \\
 1990.54$^a$ & $(1.67 \pm 0.25) \times 10^{42}$ \\
 1990.54$^a$ & $(1.52 \pm 0.23) \times 10^{41}$ \\
 1990.97$^a$ & $< 2.52 \times 10^{41}$ \\
 1991.04$^a$ & $< 2.80 \times 10^{41}$ \\
 1993.55$^a$ & $(1.97 \pm 0.29) \times 10^{40}$ \\
 1996.89$^a$ & $(1.15 \pm 0.17) \times 10^{39}$ \\
 2002.76$^b$ & $(4.95 \pm 0.99) \times 10^{39}$ \\
 2007.44$^c$ & $(4.40 \pm 0.10) \times 10^{39}$ \\
 \hline
 \end{tabular}
}
~\\
$^a$ \cite{li.etal.2002} \\ $^b$ \cite{halpern.gazari.komossa.2004}\\ $^c$ This work \\
\label{xlum} 
\end{table}

\begin{figure}[t]
\includegraphics[scale=0.4,angle=-90]{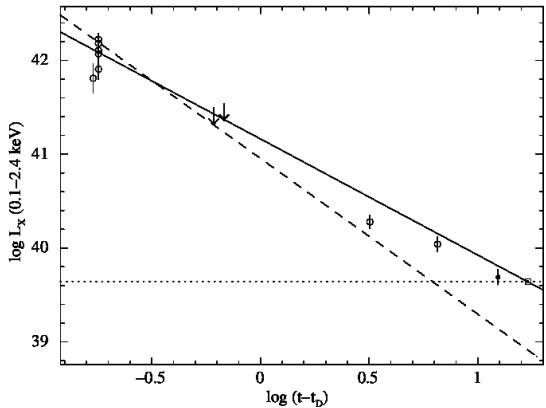}
\caption{Figure plots log$L_{\rm x}$ in the 0.1-2.4 keV energy range as a function of log$(t-t_D)$
where $t_D=1990.36$. Table \ref{xlum} gives the values of time of observation and $L_{\rm x}$
used in this plot. Data to derive these values have been taken from 
\cite{li.etal.2002} (open circle), \cite{halpern.gazari.komossa.2004} (filled circle), the 2007 luminosity being 
derived from this work and adjusted to the (0.1-2.4) keV energy band (open square). The solid line 
represents the best fit slope of $(-1.23 \pm 0.03)$; the dashed line
represents the expected (-5/3)~slope of the TDE model expected decline in 
$L_{\rm x}$ as a function of time and the dotted line represents the constant luminosity
due to nuclear star formation. 
}
\label{long-term-X-ray-lc}
\end{figure}

Figure \ref{long-term-X-ray-lc} shows a plot of the long term variation in the observed X-ray 
luminosity of the source since the 1990 X-ray flare. 
The luminosities used in this plot are listed in
table \ref{xlum}.
It is clear that the X-ray luminosity estimated from the 2002 and 2007 {\it Chandra} observations 
are comparable within errors. We note here that, for the {\it ROSAT} observations, we use the 
observed count rates given in \cite{li.etal.2002} to estimate the flux of NGC~5905 using 
WebPIMMS\footnote{http://asc.harvard.edu/toolkit/pimms.jsp} and assuming a blackbody spectrum of 
$kT = 0.06$. The declining X-ray luminosity is best fit with $L_{\rm x} \propto (t-t_D)^{\gamma}$ where 
$t_D=1990.36$, $\gamma = -1.23 \pm 0.03$ plus a constant luminosity due to nuclear star formation. 
The dashed line in figure \ref{long-term-X-ray-lc} 
is the predicted theoretical TDE curve with $\gamma=-5/3$ used to explain the 1990 X-ray flare 
of NGC~5905 \citep{li.etal.2002,halpern.gazari.komossa.2004} and the dotted 
line represents the constant luminosity due to
nuclear star formation. The $\gamma$ we get from our fitting depends on the value of $t_D$ which is fixed; hence
the real $\gamma$ value might be much closer to $-5/3$ than indicated here.

\subsection{Diffuse X-ray Emission} 
We have detected faint diffuse X-ray emission (when the bright patch 
of pixels at the optical location of NGC~5905 are subtracted) from within the bulge,
about $2^{\prime\prime}$ from the galaxy center. The AINUR H$\alpha$ image
of NGC~5905 shows the presence of a central ring of nuclear
star formation \citep{Com10}. Figure \ref{X-ray_diffuse_emission}
is an overlay of the X-ray diffuse emission contours on this H$\alpha$ map.
There is no strong spatial correlation between the H$\alpha$ and the
diffuse X-ray emission. Therefore this diffuse X-ray emission could 
be due to gas infall associated with the large scale bar in the galaxy 
and the formation of the nuclear ring.

\subsection{No AGN signatures in X-Ray, IR or Radio emission}

The X-ray image of NGC~5905 (Figure \ref{X-ray_image}) is a cluster of 
bright pixels. This cluster of bright pixels are clearly visible in 
the 0.5-2.0 keV energy band and are very faint in the 2.0-7.0 keV energy 
band. 
From the photon counts within the 3.5$^{\prime\prime}$ radius marked in figure~\ref{X-ray_image}
the hardness ratio of the source is $\sim$0.7. The estimated source luminosity in the 0.5-8.0~keV
energy band is $\sim 4.1 \times 10^{39}$ erg~s$^{-1}$. Presence of an AGN is expected if
the X-ray image shows a bright point source at the optical center with a hardness ratio greater
than 0.8 and a source luminosity greater than $3 \times 10^{42}$~erg~s$^-1$ \citep{bauer.etal.2004}.
It is also noted here that the Raymond model for emission from hot diffuse gas
gives the best fit to the X-ray spectrum indicating that the observed X-ray flux
is probably all due to the nuclear star formation.

The {\it Spitzer}/IRS mid-infrared spectrum does not show any [NeV] lines at 14.3 or $24.3\,\mu$m
which are expected if the galaxy hosts an AGN. The measured 
[OIV]$25.89\,\mu$m/[NeII]$12.81\,\mu$m ratio ($\simeq 0.02$)
indicates that the nuclear emission of this galaxy is completely
powered by star formation \citep[see][]{pereira-santaella.2010,tommasin.et.al.2010}.
To constrain the AGN contribution, if any, we can compare various line
ratios with the EW of the $11.3\,\mu{\rm m}$ PAH features using the 
diagnostic diagrams of \cite{tommasin.et.al.2010}. Accordingly we calculate 
upper limits for [NeV]14.32 $\mu$m/[NeII]12.82 $\mu$m at 0.025 and for 
[NeV]14.32 $\mu$m/[SiII]34.8 $\mu$m at 0.014. The line ratios for [OIV]/[NeII] 
and [NeIII]/[NeII] are $0.02 \pm 0.01$ and $0.12 \pm 0.02$, respectively. 
Using the estimated line ratios and the value of EW of the $11.3\,\mu{\rm m}$ 
PAH feature in Figures 4a and 4b of \cite{tommasin.et.al.2010} we conclude that
the IR flux is nearly 100\% dominated by the starburst emission.
Indeed the semi-analytical models suggested by \cite{tommasin.et.al.2010}
estimates 5.6\% AGN emission at 19~$\mu$m using the [OIV]/[NeII] line ratio 
which is consistent with the $<$34.6\% AGN emission suggested by the
upper limit on [NeV]/[NeII] line ratio.
The bright 11.3$\mu$m PAH feature and the [NeII] line
at $12.81\,\mu$m also indicate that there is high star formation 
in the nuclear region of the galaxy. From the star forming region (SFR) calibration 
given in \cite{diamond.2012} based on the [NeII] line we obtain 
a nuclear SFR (for a Salpeter IMF) of $2.3\,M_{\odot} {\rm yr}^{-1}$, 
in good agreement with the estimate of \cite{mazucca.etal.2008} 
using H$\alpha$ observations. Using the correlations between SFR 
and the soft X-ray luminosity of \cite{Pereira-Santaella.et.al.2011} 
we get an expected 0.5-2.0 keV X-ray luminosity of 
$\sim 4\times10^{39}\rm{erg~s^{-1}}$ consistent with the measured 
X-ray luminosity (see Table \ref{X-ray_spec_model}) leaving no excess 
X-ray emission which can be associated with a central Seyfert 2 type AGN.

The absence of radio emission in the 8~GHz VLA radio map also points 
to the absence of an AGN. The flux density of radio emission from AGNs 
does not fall as sharply with increasing frequency as SFRs \citep{condon.1992}. Hence if the compact radio core was due to 
an AGN, we would have observed some emission at 8~GHz as well. 

\cite{gezari.etal.2003} observed NGC~5905 in the optical using the HST/STIS 
which has a narrow slit size of $0.1^{\prime \prime}$ and hence detected 
narrow emission lines from the nuclear region of the galaxy. But they
did not detect any broad Balmer line emission. Furthermore, the spatial profile 
of the HST acquisition image obtained by them also does not indicate any AGN-like 
unresolved point source in the nucleus. Thus no evidence for an AGN 
is found from X-ray, Optical, IR or radio data. We infer that any AGN present in
NGC~5905 is very weak, as also indicated by a low total X-ray luminosity 
(see Table \ref{X-ray_spec_model}), and rule out the possibility that the 
X-ray flare of 1990 in this galaxy observed by RASS was a result of large-amplitude variability 
in the Seyfert nucleus.

\begin{figure*}[t]
\begin{minipage}{170mm}
\centering
\includegraphics[width=7cm]{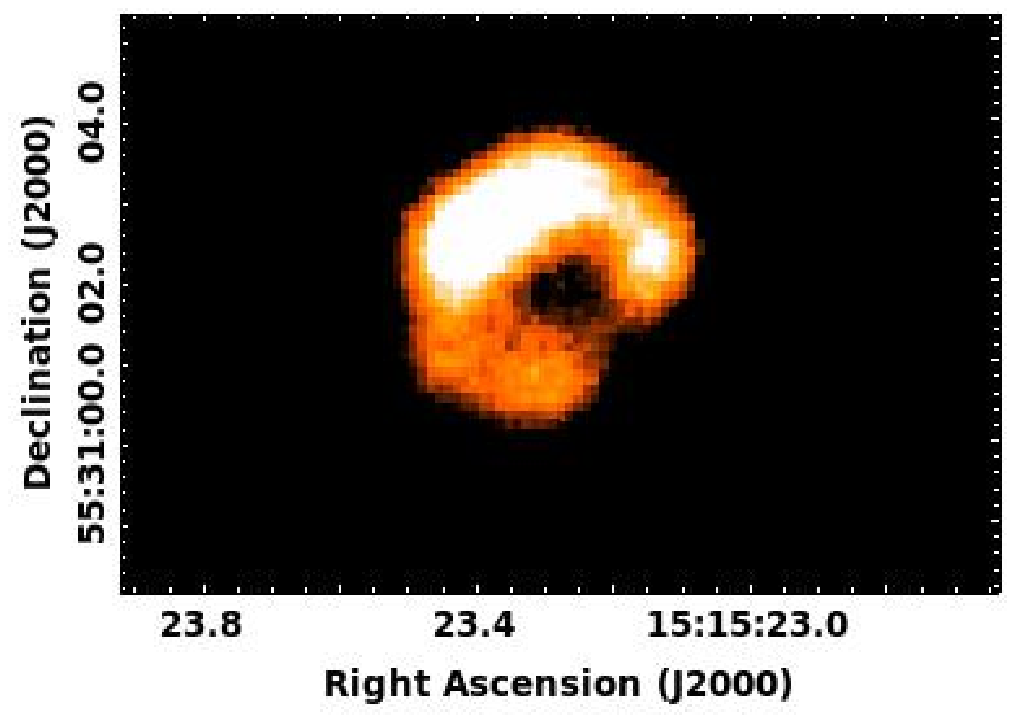}
\includegraphics[width=7cm]{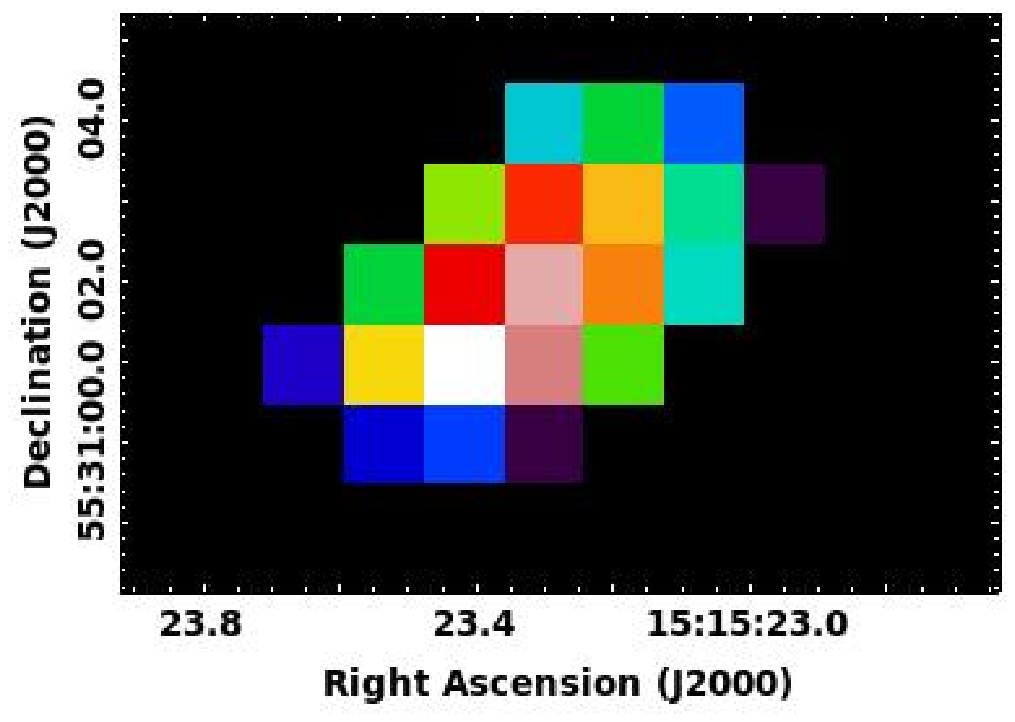}
\includegraphics[width=7cm]{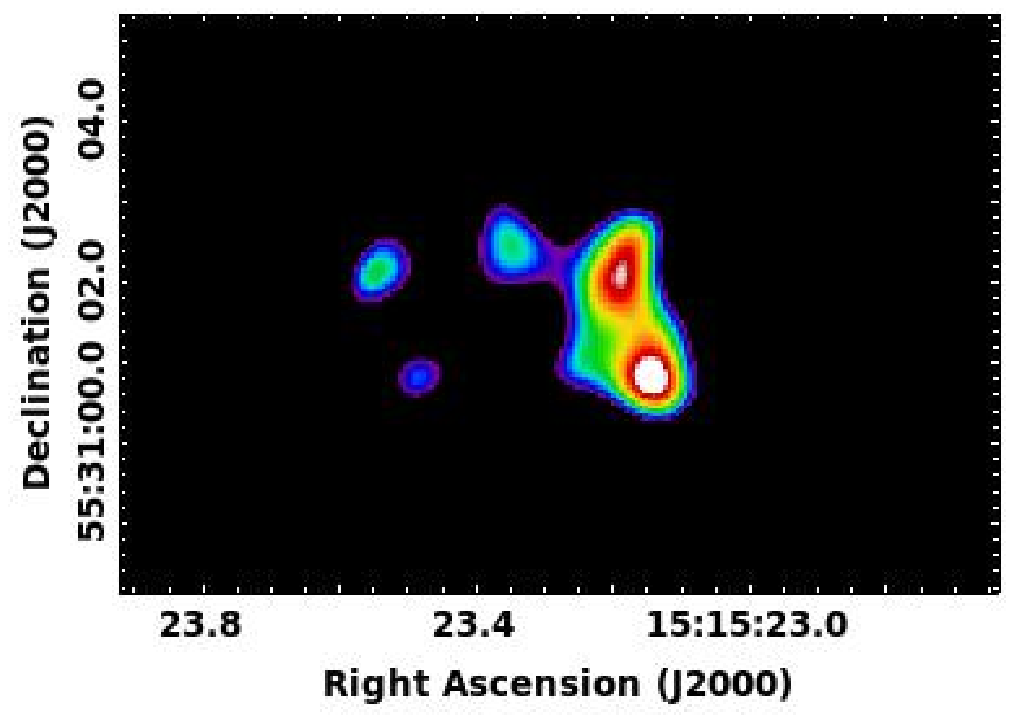}
\includegraphics[width=7cm]{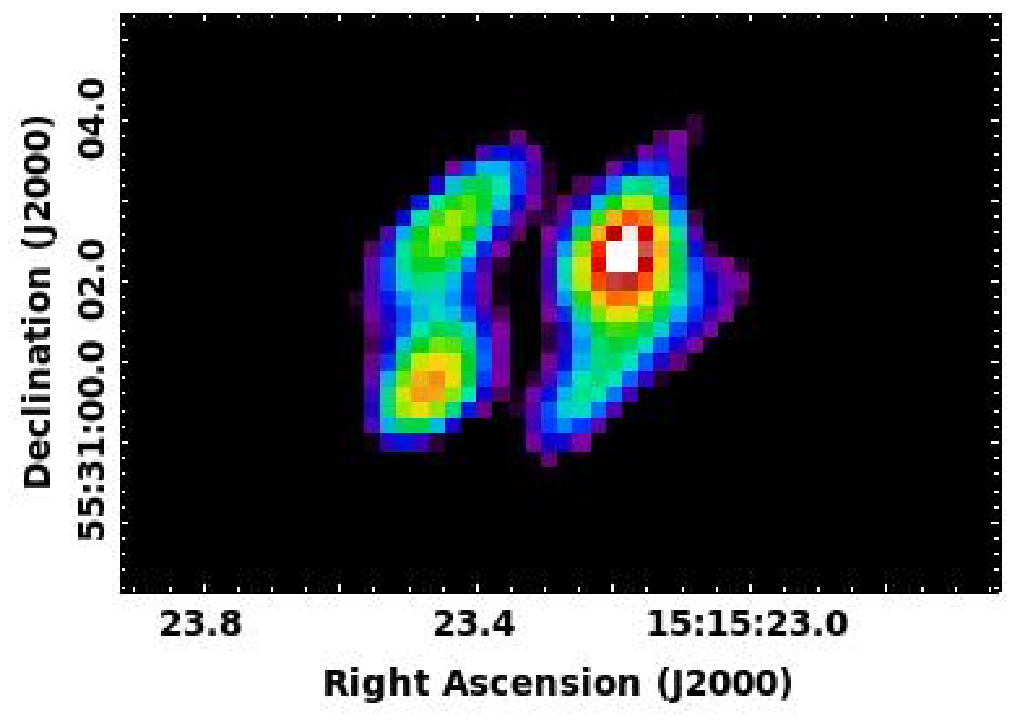}
\caption{Panel showing the inner 10$^{\prime\prime}$ nuclear star forming ring in H$\alpha$ emission and radio emission at 
1.28, 3.0 and 8.0 GHz. All the images are centered at the optical center of the galaxy (15:15:23.321, +55:31:02.51). 
Clockwise from top left (a) H$\alpha$ image from the AINUR survey (Comeron et al. 2010); (b) The 1.28~GHz GMRT  radio 
image made with robust=-5 which gives the highest resolution map. The peak flux is 1.81 mJy/beam where the beam 
is 4.97$^{\prime\prime} \times$1.63$^{\prime\prime}$ and the map noise is  87.6~$\mu$Jy/beam. (c) the 3~GHz VLA image 
made with robust=-5. The peak flux 
is 0.74~mJy/beam where the beam is 1.84$^{\prime\prime} \times$1.17$^{\prime\prime}$ and the map noise is  
81.45~$\mu$Jy/beam. Note that the center of 
the circum nuclear ring coincides with the location of the peak emission in the GMRT 1.28~GHz map. (d) 
The VLA 8~GHz map made with robust=0 (as it has more emission than robust=-5) and convolved with a 1$^{\prime\prime}$ beam to 
obtain a smoothed map. Note that the emission appears to be associated with the circum nuclear ring.} 
\label{panel}
\end{minipage}
\end{figure*}

\subsection{Radio emission at 1.28~GHz and 3~GHz~:~the circumnuclear star formation}
At all resolutions in the GMRT maps, the 1.28~GHz radio emission
shows strong nuclear emission but no extended structure that could be associated with large scale jets or 
disk emission (see Figure~\ref{radio_emission}). The source structure and flux density are 
similar to that of the FIRST image (Table~\ref{radio_obs_detail}).
The NVSS map has a flux density of 20.1~mJy which is significantly higher  
than both the FIRST and GMRT maps. But this may be due to the larger NVSS beam 
(45$^{\prime\prime}$) which picks up more of the extended diffuse emission.
To see if there is any compact nuclear emission in the 1.28~GHz radio image, we made a high resolution
map using the imaging parameter, robust=-5 (Figure~\ref{panel}(b)). The emission peaks close to the galaxy nucleus and
has a value of 1.81 mJy/beam where the beam is 4.97$^{\prime\prime} \times$1.63$^{\prime\prime}$ and 
the map noise is 0.088~mJy/beam. But it is offset
from the optical center by approximately an arcsecond. The emission is also extended. Hence we could not detect
any compact emission that could be a radio afterglow associated with the TDE at 1.28~GHz.

At 3~GHz, however, the 1.28~GHz nuclear emission breaks up into a double lobed structure
(Figure~\ref{panel}(c)) that could be either due to a weak AGN or emission 
from the limb brightened edges of a nuclear ring. Since there is a circumnuclear star 
forming ring in the galaxy ((Figure~\ref{panel}(a)), a nuclear ring is more likely. The size and orientation
of the ring matches the high resolution 1.28~GHz radio morphology. The size is also similar to the H$\alpha$
image of the circumnuclear ring. All these factors suggest the 3~GHz emission is mainly due to circumnuclear 
star formation. Using the total flux at 1.28 and 3~GHz (Table~\ref{radio_obs_detail}) we obtain a spectral index of
$\alpha~=~0.5$ where $S_{\nu}\sim{\nu}^{-\alpha}$. This is typical for both radio jets and 
star formation. However the high resolution radio geometry suggests that star formation is more likely. 

At 8~GHz we obtained only patchy emission. To see if there was any extended emission that was not picked up by the
small beam, we smoothed the 8~GHz map (robust=0, lower resolution) to 1$^{\prime\prime}$ 
resolution (Figure~\ref{panel}(d)). No compact
emission that could be due to a TDE was detected, but some patchy emission close to the noise limit is seen.
It is probably associated with the star forming ring.

We used the radio flux at 1.4~GHz in the FIRST map (Table~\ref{radio_obs_detail}) to estimate the star formation rate in 
the nuclear region using the relation $\rm{SFR}=(L_{1.4}/4\times10^{21}~WHz^{-1})~M_{\odot}yr^{-1}$\citep{condon.1992}. 
We obtain a value of 0.61~${\rm M_{\odot}yr^{-1}}$, which is lower than the value
that we obtained from the {\it Spitzer} spectrum, 2.3~$\rm{M_{\odot}yr^{-1}}$ and from the H$\alpha$
observations of the nuclear ring, 2.6~$\rm{M_{\odot}yr^{-1}}$ \citep{mazucca.etal.2008}. This could be 
because the SFR derived from radio continuum assumes the presence of massive 
stars which are usually associated with supernova that emit synchrotron radiation. However, NGC~5905 is a 
giant LSB galaxy and in such systems massive stars may not be common. 

One question that arises is whether the TDE could have triggered the nuclear star formation through gas
shocked by radio jets. This is
unlikely as the ring has an inner diameter of 0.5$^{\prime\prime}$ to 1$^{\prime\prime}$. Even with velocity 
close to speed of light ($c$), 
the time required to reach such radius is of the order of several $\sim100$~yrs. However, the
higher nuclear mass density makes a TDE more probable.

\subsection{Nuclear radio emission~:~Is there a TDE radio afterglow?}  

As mentioned in the introduction, TDEs can result in enhanced 
SMBH mass accretion rates which can result in the formation of radio jets or 
outflows from the nucleus. They will appear as relatively weak but
compact radio emission associated with the nucleus. One of the main goals of this study 
is to detect such a radio afterglow from NGC~5905. To search for compact emission we
applied 20-0 and 40-0 k$\lambda$ UV range cutoffs (i.e. only UV 
wavelengths greater than 20 or 40 k$\lambda$ were included) to both the 1.28~GHz
and 3~GHz maps (see Table~\ref{radio_obs_detail} for fluxes). The 1.28~GHz 20~k$\lambda$ map
has a peak coincident with the center
of the 3~GHz ring (Figure~\ref{panel}). But when a 40~k$\lambda$ cutoff
is applied, no compact emission remained - which suggests
that the 20~k$\lambda$ 1.28~GHz peak was probably due to diffuse emission associated
with star formation and not compact emission associated with the TDE.  
The 3~GHz images with 20-0 and 40-0~k$\lambda$ UV range cutoffs
(Figure~\ref{panel} b and c) show a nice ring but no compact nuclear emission. 
Thus both the 1.28 and 3~GHz radio data (both obtained after the epoch 2010) do not show any emission that
could be associated with the TDE. But the map noise at 1.28~GHz and 3~GHz constrains the
radio afterglow to lie between 0.62~mJy and 0.09~mJy (3$\sigma$ limits), respectively. We did not 
detect any emission in the VLA 8~GHz (1996) archival data  as well. 
Instead the map noise gives a 5$\sigma$ limit of 0.15~mJy for an 8~GHz afterglow.

\section{Conclusion }
\noindent
{\bf 1. X-ray emission:} From the 2007 {\it Chandra} X-ray spectrum of NGC~5905 we conclude that the observed X-ray emission
at that epoch can be fully accounted for as emission from a hot diffuse gas probably present due
to the nuclear star formation. \\
{\bf 2. No evidence for an AGN:} No evidence for an AGN is found from X-ray, IR or radio observations. 
The predicted 0.5-2keV X-ray luminosity obtained from the IR SFRs compares well with the observed  X-ray luminosity. 
The non-detection of the high excitation [NeV] lines together with the observed EW of the 11.3~$\mu$m PAH feature and 
mid-IR line ratios indicate that the mid-IR emission of this galaxy is fully arising from star formation 
activity. According to the \cite{tommasin.et.al.2010} semi-analytical models, the [OIV]/[NeII] line ratio 
suggests an AGN contribution of only 5.6\% to the total IR flux at 19~$\mu$m. Also radio emission at 8~GHz is absent. 
Therefore the STIS data \citep{gezari.etal.2003} can imply only a very weak AGN in NGC~5905. \\
{\bf 3. Circumnuclear star formation:} The 3~GHz image of the galaxy shows a double lobed structure the size of
which is similar to the H$\alpha$ image of the circumnuclear ring. The size and orientation of this structure
matches with the 20~k$\lambda$ 1.28~GHz radio morphology. Thus the 3~GHz radio emission is likely due to the circumnuclear 
star forming ring in NGC~5905. \\
{\bf 4.~No Confirmed Radio Afterglow~:~} We did not detect any compact emission associated with the TDE at
1.28~GHz, 3~GHz or 8~GHz. Such emission could arise from a radio jets triggered by the TDE.  Upper
limits to any flux associated with a radio afterglow can be given by 3$\sigma$ (where $\sigma$ is the map
noise) and is 0.62~mJy at 1.28~GHz, 0.09~mJy at 3~GHz and 0.09~mJy at 8~GHz.

{\bf 5. Long term X-ray luminosity:} The long term X-ray luminosity shows a decline after the 1990 X-ray flare 
detected by RASS. This decline follows the expected decline if the 1990 X-ray flare was due to TDE. The
observed luminosity decline is well modeled as $L_{\rm x} \propto (t-t_D)^{\gamma}$ where $t_D$ is the year 1990.36, 
$\gamma = -1.23 \pm 0.03$ plus a constant luminosity due to nuclear star formation.

\section*{Acknowledgments}
We thank Sebastien Comeron for giving us the AINUR image of the nuclear ring in NGC~5905,
M. Pereira-Santaella for his help with the {\it Spitzer}/IRS spectra, S. Tommasin for help
with their semi-analytical models of IR emission lines and S.~Komossa for very useful 
discussions regarding this work. We also thank G.C. Bower for his very useful comments 
which have improved this work and for sharing the 3~GHz EVLA data used in the work.
A.A.-H. is party funded by the Spanish Plan National grant AYA2012-31447.
We thank the GMRT staff for help in observations. 
The GMRT is operated by the National Center for Radio Astrophysics of
the Tata Institute of Fundamental Research. This work has used VLA 3~GHz and 8~GHz archival data of
NGC~5905. It has also used the NRAO VLA FIRST image and the 8~GHz NRAO VLA Archival data of NGC~5905.
The NRAO (National Radio Astronomy Observatory) is a facility of the National Science Foundation operated 
under cooperative agreement by Associated Universities, Inc.
This research has made use of data obtained from the 
{\it Chandra} Data Archive and the {\it Chandra} Source Catalog, and software provided by the 
{\it Chandra} X-ray Center (CXC) in the application package CIAO. This work is based [in part] on 
observations made with the {\it Spitzer} Space Telescope, which is operated by the Jet Propulsion Laboratory, 
California Institute of Technology under a contract with NASA.
This research has also made use of the NASA/IPAC Extragalactic Database (NED) which is operated by the
Jet Propulsion Laboratory, California Institute of Technology, under contract with the
National Aeronautics and Space Administration. 




\bibliographystyle{spr-mp-nameyear-cnd}
\bibliography{ms}
\end{document}